\documentclass{article}
\usepackage{spconf,amsmath,graphicx}
\usepackage{amssymb,amsmath}
\usepackage{cite}
\usepackage{xcolor}

\title{A practical distributed active noise control algorithm overcoming communication restrictions}

\name{Junwei Ji, Dongyuan Shi, Zhengding Luo, Xiaoyi Shen, Woon-Seng Gan}
\address{School of Electrical and Electronic Engineering, Nanyang Technological University, Singapore.\\Email: JUNWEI002@e.ntu.edu.sg; dongyuan.shi@ntu.edu.sg}
\begin{document}
\ninept
\maketitle
\begin{abstract}
By assigning the massive computing tasks of the traditional multichannel active noise control (MCANC) system to several distributed control nodes, distributed multichannel active noise control (DMCANC) techniques have become effective global noise reduction solutions with low computational costs. However, existing DMCANC algorithms simply complete the distribution of traditional centralized algorithms by combining neighbour nodes' information but rarely consider the degraded control performance and system stability of distributed units caused by delays and interruptions in communication. Hence, this paper develops a novel DMCANC algorithm that utilizes the compensation filters and neighbour nodes' information to counterbalance the cross-talk effect between channels while maintaining independent weight updating. Since the neighbours' information required barely affects the local control filter updating in each node, this approach can tolerate communication delay and interruption to some extent. Numerical simulations demonstrate that the proposed algorithm can achieve satisfactory noise reduction performance and high robustness to real-world communication challenges. 
\end{abstract}
\section{Introduction}
\label{sec:intro}

In recent years, the issue of noise has attracted more attention because it not only affects people's quality of life but also causes health problems \cite{RN25}. The active noise control (ANC) system, which generates anti-noise from a secondary source to suppress unwanted noise \cite{KuoANC1999}, is one of the techniques that can assist us in resolving these issues. It has achieved significant success  with numerous commercial products. Nowadays, global noise control over a large spatial area with multiple secondary sources and error sensors is of great interest \cite{ElliottMEANC1987}. Due to its greater degree of control freedom, the multichannel active noise control (MCANC) system is widely used in this scenario to expand the quiet zone \cite{CheerVehicles2015,MuraoWindow2017,ElliottWave2018,IwaiSepar2019,WangCavity2019}.

One of the most widely used algorithms for MCANC system is the multichannel filtered X least mean square (McFXLMS) algorithm \cite{ElliottMEANC1987}, based on which control strategies are categorized as either centralized or decentralized methods. In a centralized system, the controller will process all input and output signals, necessitating a significant amount of processing power~\cite{shi2019practical}. Consequently, small-scale MCANC systems typically employ a centralized approach. To address this issue, the decentralized approach is proposed to allocate the entire computation tasks to different processors at the cost of global reduction performance \cite{LebDecen2002,ZhangPerDecen2013,ZhangDecen2019,AnOptidecen2021}. For the purpose of further enhancing the performance of the decentralized structure, a distributed multichannel active noise control (DMCANC) system with an incremental strategy \cite{LopesIncrem2007} is developed to achieve the same steady-state performance as the centralized structure under an ideal network\cite{FerrerInANC2015}. However, the limited communication speed and computational burden at each node hinder the massive deployment of the technology in practice. In contrast, the diffusion multichannel ANC system, where each node only collaborates with its neighbours \cite{LopesDiff2008,ChenDiffMultask2015}, reduces the computational burden \cite{AntoDiff2015,SongDiff2016,ChuDiffICSV2017,KukdeDiff2019}. It is demonstrated that the diffusion MCANC system can achieve a similar steady-state noise reduction performance regarding the broadband and narrowband noise with a lower computational load when compared to the centralized one \cite{ChuDiffanly2020,ChenDiffnarrow2022}.
\begin{figure}[!t]
    \centering
    \includegraphics[width = 8cm]{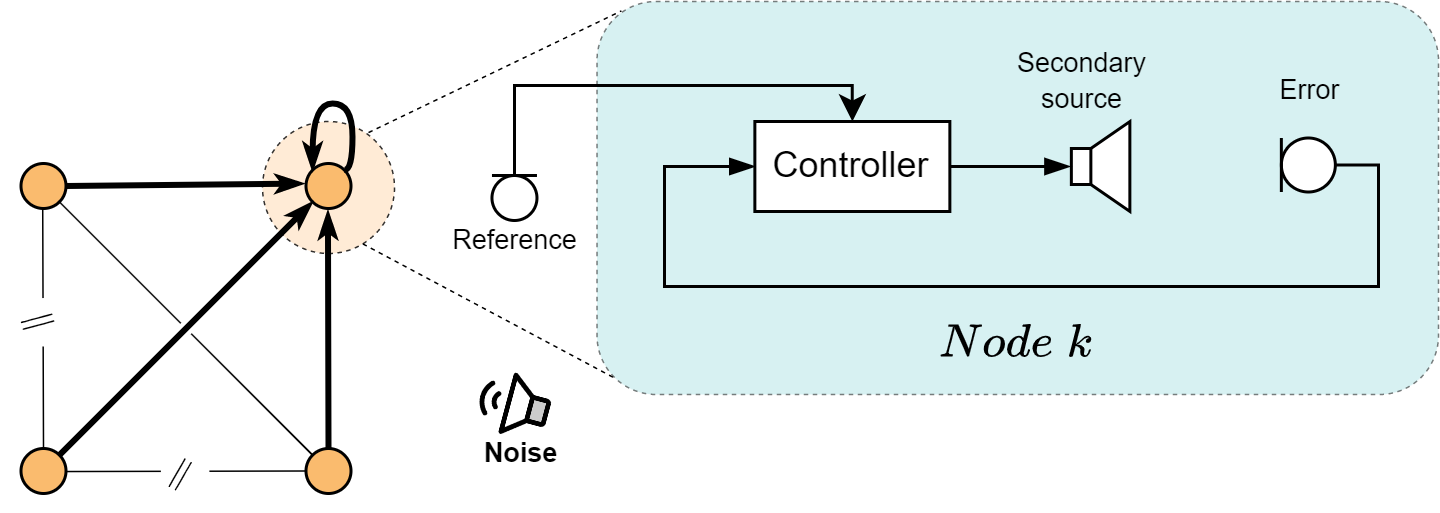}
    \caption{Distributed ANC network with nodes consisting of a single controller, a single secondary source, and a single error sensor}
    \label{fig:1 node}
\end{figure}

Up to now, most DMCANC utilizes topology-based combination rules \cite{XiaoDistr2004,ScherberDistri2004,BlondelDistr2005,LinDistr2005} to integrate received data, ignoring the cross-talk effect among nodes. To address this issue, we incorporate compensation filters into the system to circumvent this issue. By employing the compensation filters rather than the combination rule to combine the local control filters from other nodes, the proposed DMCANC mitigates the inter-node cross-talk to a certain extent. Additionally, it only updates the local control filters for each node based on the local error signal. Furthermore, as far as we are aware, the majority of DMCANC system assumes that all the nodes can exchange information with other nodes at the same time \cite{ChenDiffnarrow2022} without the consideration of communication restrictions \cite{LeeDiscommdelay2015,ZhaoDiscomminterr2022}, which is unrealistic. In this paper, the communication delays and interruptions are taken into account, rendering the DMCANC system more applicable. The simulation results show that the proposed method not only has the equivalent steady-state noise reduction performance as the centralized MCANC but also exhibits great robustness to communication restrictions.


The remaining sections of this paper are structured as follows: Section \ref{sec:proposed} derives the compensation filters and describes the proposed DMCANC algorithm. In Section \ref{sec:simulation}, several numerical simulations are conducted to illustrate the proposed method's validity and robustness to the communication restrictions. Finally, conclusions are drawn in Section \ref{sec:conclusion}.
 
\section{Methodology}
\label{sec:proposed}

This section aims to describe the modified DMCANC algorithm in that the compensation filters, which will be explained in Section \ref{ssec:CompenFil}, are developed to substitute the conventional combination rule. As a result, the DMCANC algorithm needs to be improved owing to extra disturbance introduced by compensation filters, which is discussed in Section \ref{ssec:newDMCANC}.

\subsection{Compensation Filters}
\label{ssec:CompenFil}
\begin{figure}[!t]
    \centering
    \includegraphics[width = 8.5cm]{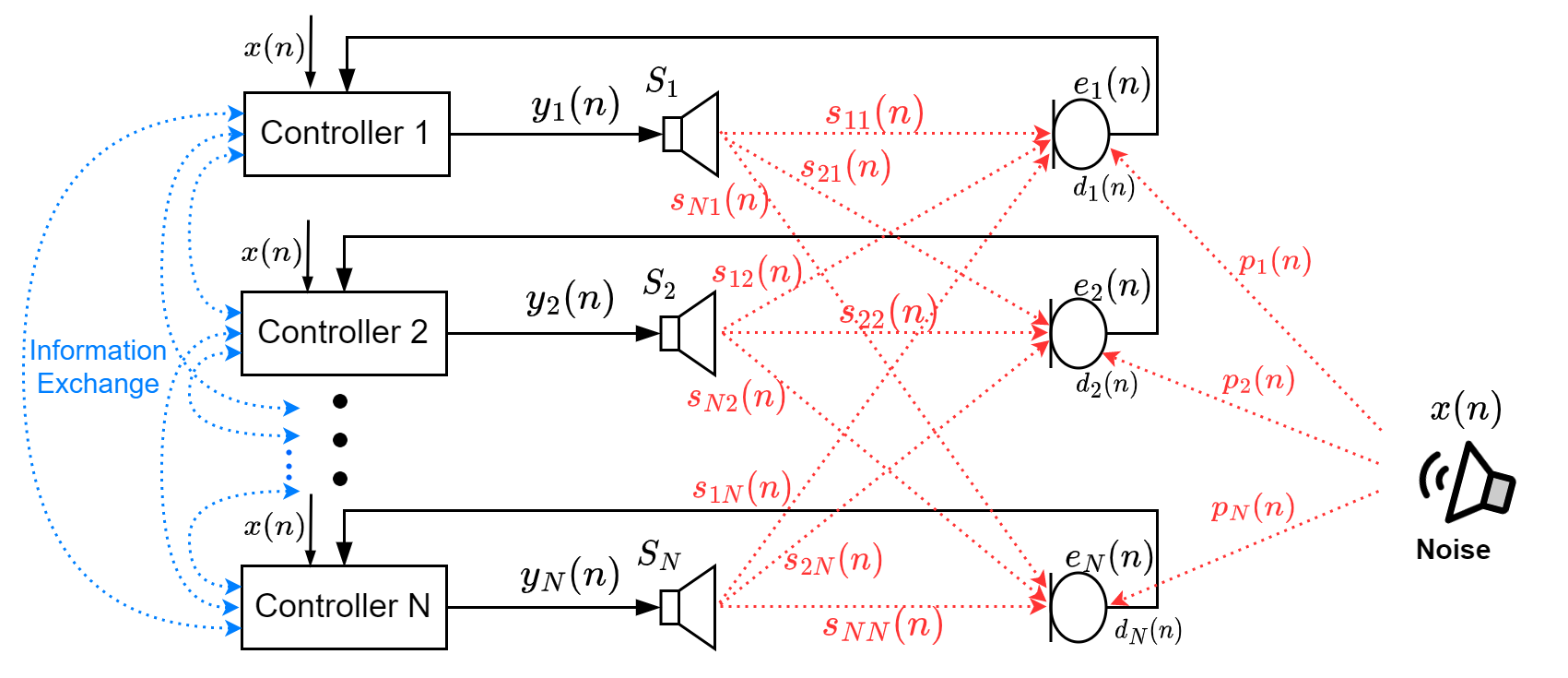}
    \caption{The schematic diagram of DMCANC network}
    \label{fig:2 DMCANC}
\end{figure}

As depicted in Fig.~\ref{fig:1 node}, a DMCANC system typically consists of multiple nodes, each of which consists of a single controller, a single secondary source, and a single error sensor. Each node can independently execute the same algorithm and collaborate with other nodes via communication to meet the system's overall performance requirements.

Figure~\ref{fig:2 DMCANC} illustrates a DMCANC network with $N$ nodes. In terms of a particular node, such as the $k$th node, it captures the reference signal $x(n)$ and generates the control signal $y_k(n)$, which travels along the secondary path to attenuate disturbances. Here $n$ denotes the time instant and the secondary path refers to the acoustic path from the secondary source to the error sensor. Thus, the disturbance $d_k(n)$ on the $k$th node is cancelled by the anti-noises originating from all secondary sources, and its residual error $e_k(n)$ is measured by the $k$th error sensor as:
\begin{equation}\label{eq1}
    e_{k}(n) = d_{k}(n) - \sum_{m=1}^{N}{y}_{m}(n)*{s}_{km}(n),
\end{equation}
where $*$ represents the convolution operation, and ${s}_{km}(n)$ denotes the secondary path from $m$th secondary source to the $k$th error sensor. 

However, the control signal $y_m(n),(m\neq k)$ received at the $k$th node can be regarded as interference to that node. Hence, $e_k(n)$ can be rewritten as
\begin{equation}\label{eq2}
    e_{k}(n) = d_{k}(n) -{y}_{k}(n)*{s}_{kk}(n) - \gamma_{k}(n),
\end{equation}
where the interference $\gamma_{k}(n)$ is given by
\begin{equation}\label{eq3}
    \gamma_{k}(n) = \sum_{m=1,m\neq k}^{N}{y}_{m}(n)*{s}_{km}(n).
\end{equation}
 
A straightforward notion is to use the $k$th node to generate the sound wave to counterbalance this interference $\gamma_k(n)$ based on the information from the other nodes, such as their control filters' coefficients. Due to the fact that these nodes have distinct secondary paths, compensation filters, ${c}_{km}(n)$ are introduced to help the $k$th node eliminate interference from other secondary sources. The relationship between this compensation filter and other secondary paths is expressed as:
\begin{equation}\label{eq4}
    {s}_{km}(n) = {s}_{kk}(n)*{c}_{km}(n), m\neq k.
\end{equation}
Here, we name ${s}_{kk}(n)$ as the self-secondary path and ${s}_{km}(n), (m\neq k)$ as the cross-secondary path.

\begin{figure}[!t]
    \centering
    \includegraphics[width = 8cm]{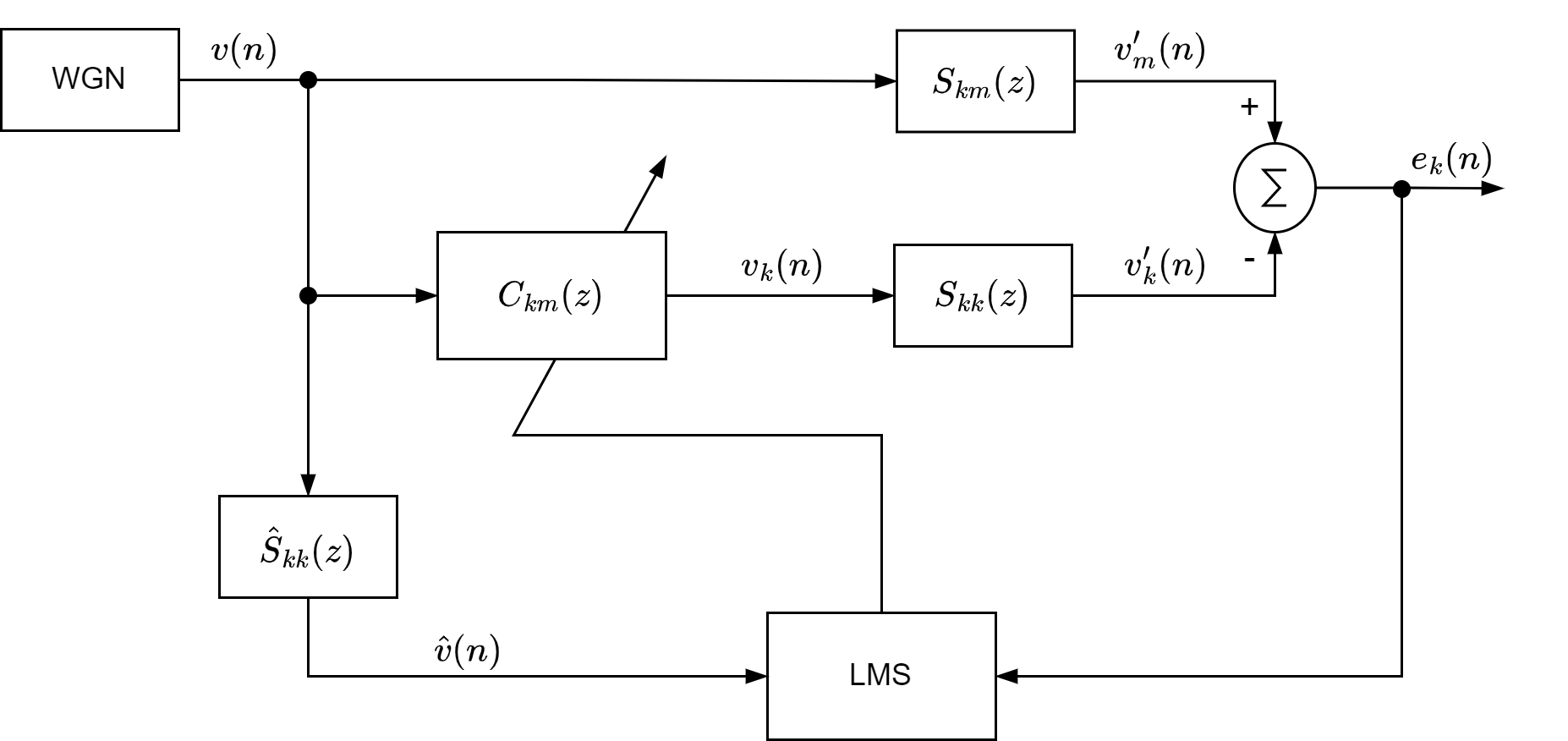}
    \caption{The block diagram of obtaining the compensation filter using the FXLMS algorithm.}
    \label{fig:3 SPCompensation}
\end{figure}

To obtain the compensation filter, the FXLMS algorithm is used \cite{KuoANC1999}, whose block diagram is shown in Fig.~\ref{fig:3 SPCompensation}. The $S_{km}(z)$, $S_{kk}(z)$ and $C_{km}(z)$ are the transfer function of ${s}_{km}(n)$, ${s}_{kk}(n)$ and ${c}_{km}(n)$ respectively.
In this figure, the White Gaussian Noise (WGN), $v(n)$, passes through the cross secondary path $S_{km}(z)$ and forms the desired signal $v^\prime_m(n)$. The $C_{km}(z)$ is used to compensate for the difference between $S_{km}(z)$and $S_{kk}(z)$, and its impulse response can be derived recursively using the FxLMS algorithm as:

\begin{equation}\label{eq5}
    \mathbf{c}_{km}(n+1) = \mathbf{c}_{km}(n) + \mu_{c}\mathbf{\hat{v}}(n)e_k(n), m\neq k,
\end{equation}
where $\mu_{c}$ is the step size for compensation filter update and the vector $\hat{\mathbf{v}}(n)$ is the input vector $\mathbf{v}(n)$ filtered by estimated secondary path $\hat{S}_{kk}(z)$. The $e_k(n)$ in \eqref{eq5} is expressed as:
\begin{equation}\label{eq6}
    e_{k}(n) = v'_{m}(n)-v'_{k}(n),
\end{equation}
where $v'_k(n)$ is formed by passing the exciting WGN signal through the compensation filter and the self-secondary path.

\subsection{Proposed DMCANC}
\label{ssec:newDMCANC}
After estimating the compensation filters node by node through offline modelling, we will use them to combine the local control filters of other nodes with that of the current node in order to generate the control signal. This combined control filter is called the global control filter ${w}_{k}(n)$, which is defined as:
\begin{equation}\label{eq7}
    {w}_{k}(n) = {\psi}_{k}(n)-\sum_{m=1,m\neq k}^{N}{\psi}_{m}(n)*{c}_{km}(n),
\end{equation}
where ${\psi}_{k}(n)$ is the local control filter of node $k$, and its control signal $y_k(n)$ is obtained from
\begin{equation}\label{eq8}
    y_k(n)={x}(n)\ast {w}_{k}(n),
\end{equation}
where $x(n)$ is the reference signal. According to (\ref{eq4}), (\ref{eq7}) and (\ref{eq8}), (\ref{eq2}) can be derived to 

\begin{equation}\label{eq9}
\begin{split}   
    e_k(n) & = d_{k}(n) -x(n)*{\psi}_{k}(n)*{s}_{kk}(n) \\
     & +\sum_{m=1,m\neq k}^{N}{x}(n)*{\psi}_{m}(n)*{s}_{km}(n)- \gamma_{k}(n).
\end{split}
\end{equation}

Meanwhile, based on \eqref{eq7} and \eqref{eq8}, the interference $\gamma_{k}(n)$ defined in \eqref{eq3} can be rewritten as:
\begin{figure}[!t]
    \centering
    \includegraphics[width = 8cm]{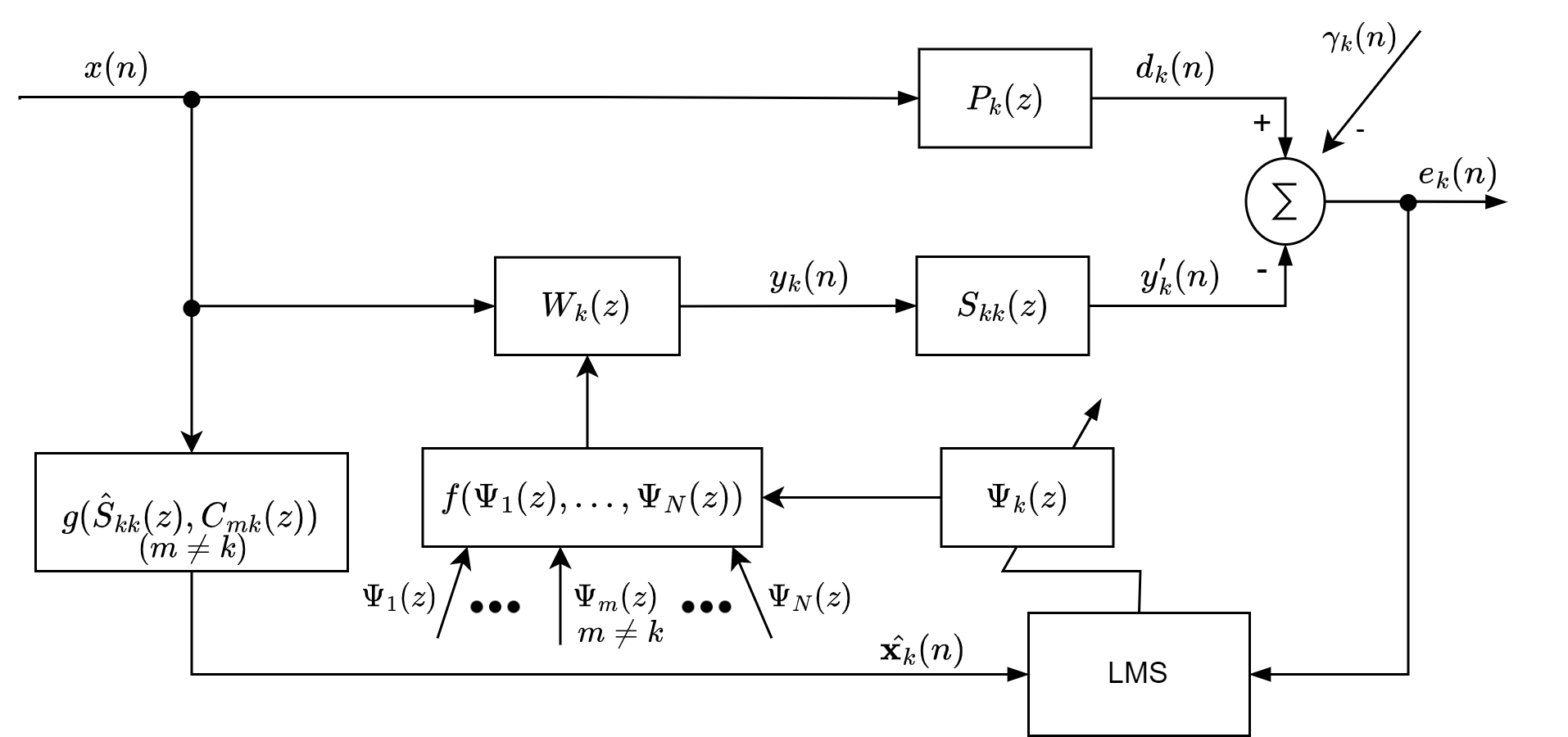}
    \caption{Block diagram of proposed DMCANC for the $k$th node. $\Psi_{k}(z)$ is the transfer function of ${\psi}_{k}(n)$. $f(\Psi_{1}(z),...,\Psi_{N}(z))$ and $g(\hat{S}_{kk}(z),C_{mk}(z))$ are defined in (\ref{eq7}) and (\ref{eq14}) respectively}
    \label{fig:4 NodekDMCANC}
\end{figure}
\begin{equation}\label{eq10}
\begin{split}
    \gamma_{k}(n) &= \sum_{m=1,m\neq k}^{N}{x}(n)*{\psi}_{m}(n)*{s}_{km}(n)\\
    & -  {x}(n)*\sum_{m=1,m\neq k}^{N}[{\psi}_{k}(n)*{c}_{mk}(n) \\
    & - \sum_{l=1,l\neq k,m}^{N}{\psi}_{l}(n)*{c}_{ml}(n)]*{s}_{km}(n).
\end{split}
\end{equation}
Furthermore, substituting (\ref{eq4}) and (\ref{eq10}) into (\ref{eq9}) yields
\begin{equation}\label{eq11}
\begin{split}
    e_k(n) & = d_{k}(n)-{x}(n)*\\
    &\sum_{m=1,m\neq k}^{N}\sum_{l=1,l\neq k,m}^{N}{\psi}_{l}(n)*{c}_{ml}(n)*{s}_{km}(n)\\
    &-{x}(n)*[{\psi}_{k}(n)-\\
    & \sum_{m=1,m\neq k}^{N}{\psi}_{k}(n)*{c}_{mk}(n)*{c}_{km}(n)]*{s}_{kk}(n).
\end{split}
\end{equation}
Since cross-talk between different nodes has been reduced by compensation filters, global control of the system can be attained by minimizing the sound pressure levels of the error sensor using only the controller of the corresponding node. Hence, the local cost function for the $k$th node is defined as
\begin{equation}\label{eq12}
  J_{local\_k} = e_k^2(n).
\end{equation}
In order to minimize \eqref{eq12}, its negative gradient in terms of $\boldsymbol{\psi}_{k}(n)$ is used to update the local control filter vector as:
\begin{equation}\label{eq13}
  \boldsymbol{\psi}_{k}(n+1) = \boldsymbol{\psi}_{k}(n)+\mu_{\psi}\hat{\mathbf{x}}_{k}(n)e_{k}(n),
\end{equation}
where $\mu_{\psi}$ is the step size for the local control filter update. $\hat{\mathbf{x}}_{k}(n)=[\hat{{x}}_{k}(n),...,\hat{{x}}_{k}(n-L_{\psi}+1)]$ is the filtered reference vector in which $L_{\psi}$ is the length of local control filters and
\begin{equation}\label{eq14}
\begin{split}
  \hat{{x}}_{k}(n)=&{x}(n)*{s}_{kk}(n)\\
  &-{x}(n)*{s}_{kk}(n)*\sum_{m=1,m\neq k}^{N}{c}_{mk}(n)*{c}_{km}(n).
\end{split}
\end{equation}

In the proposed DMCANC, each node only updates its own local control filter based on (\ref{eq12}) and (\ref{eq13}), while those received from other nodes remain unchanged. These local control filters are then combined into a global control filter with the compensation filters which are used to make up the difference between self and cross-secondary paths. Therefore, the proposed DMCANC system can effectively reduce external noise, while suppressing inter-node cross-talk.

\section{Simulation Result}
\label{sec:simulation}
In this section, we validated the proposed method on a four-node distributed ANC system through numerical simulation. The primary noise is a broadband signal with a frequency range of $100$ to $1000$ Hz and the sampling frequency is 16000Hz. The primary and secondary paths are all band-pass filters with a frequency band between $50$ and $5000$ Hz. To simulate the general case in which the cross-secondary paths are longer than the self-secondary path for each node, the tap length of the cross and self-secondary paths are set to 320 and 256 respectively. For the convenience of the simulation, the self-secondary path length is extended to 320 with zero padding. In addition, the control filters and compensation filters consist of $512$ and $64$ taps, respectively, and the step size of algorithms is $1\times10^{-5}$. Comparative analyses between the centralized ANC approach and the proposed algorithm and the performance of the proposed method under communication constraints are discussed in the following simulations. All the mean square error (MSE) results are averaged over 30 runs.

\begin{figure}[!t]
    \centering
    \includegraphics[width = 8cm,height=6cm]{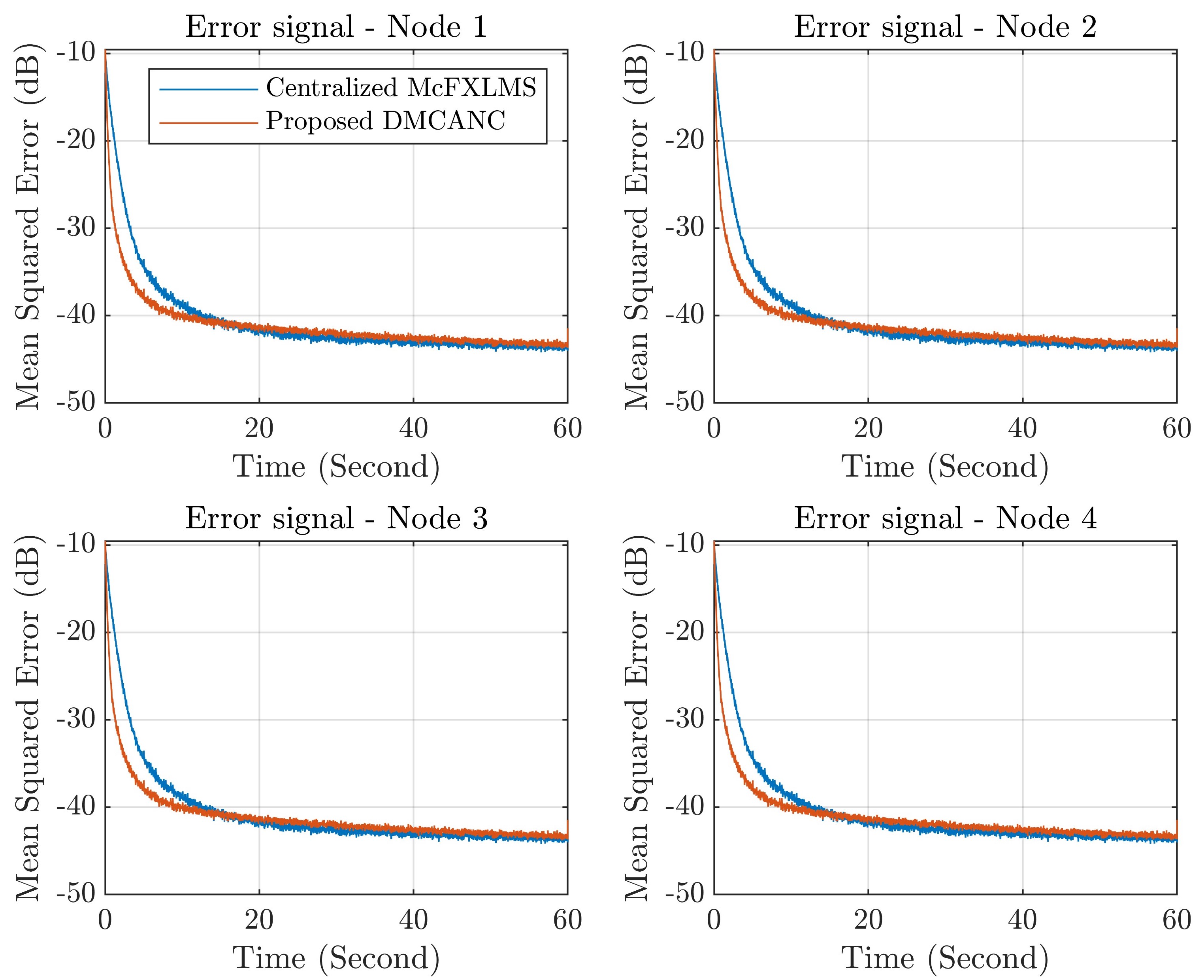}
    \caption{Mean Squared Error (MSE) of the four-node ANC network using the centralized McFXLMS and proposed DMCANC algorithm}
    \label{fig:5 Case1MSE}
\end{figure}
\begin{figure}[!t]
    \centering
    \includegraphics[width = 8cm,height=6cm]{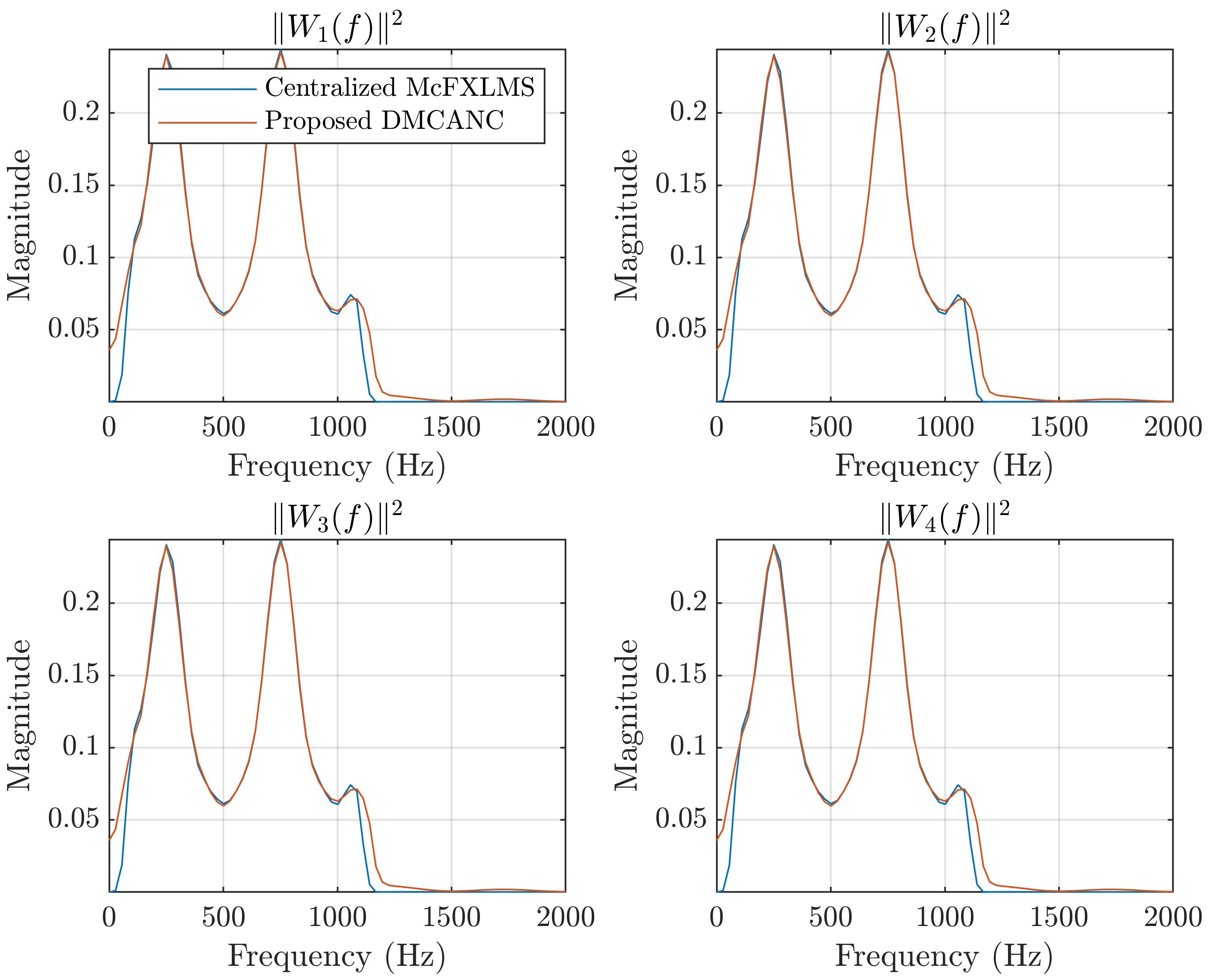}
    \caption{Frequency spectrum of the global control filters in the centralized McFXLMS and proposed DMCANC algorithm}
    \label{fig:6 Case1Weight}
\end{figure}

{\it Simulation 1:} In the first simulation, we compared the proposed DMCANC algorithm with the centralized McFXLMS algorithm \cite{ElliottMEANC1987} under the same configurations, which have been mentioned before. From the results obtained in Fig.~\ref{fig:5 Case1MSE}, it can be observed that the proposed algorithm can achieve steady-state noise reduction performance comparable to the centralized McFxLMS. It is further shown in Fig.~\ref{fig:6 Case1Weight} that these two algorithms have identical global control filters in steady-state. Therefore, the noise cancellation performance of the proposed DMCANC is not affected.

\begin{figure}[!t]
    \centering
    \includegraphics[width = 8cm,height=6cm]{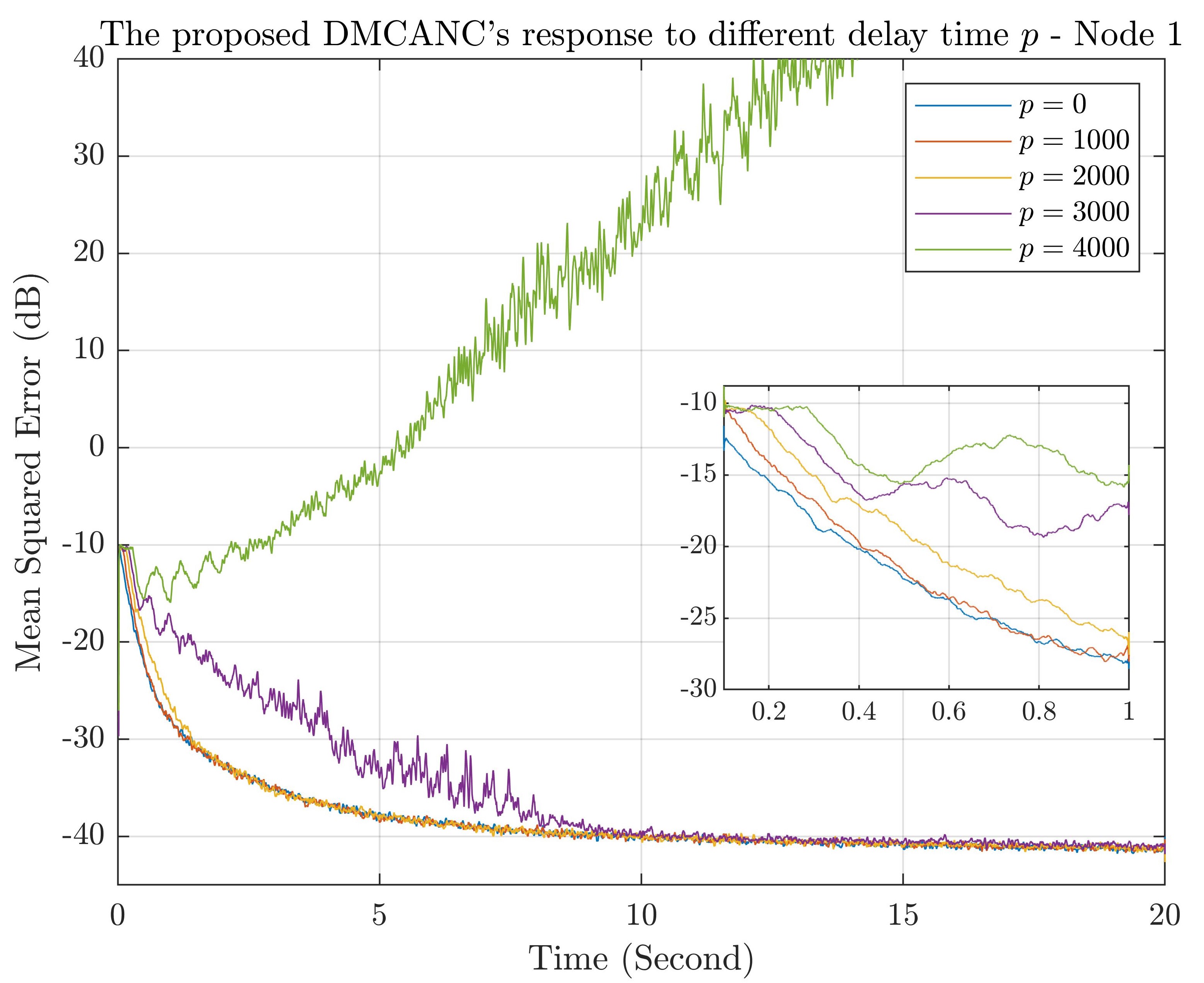}
    \caption{Mean Squared Error (MSE) of the proposed DMCANC with response to different delays}
    \label{fig:7 Case2MSE}
\end{figure}

{\it Simulation 2:} In previous discussions \cite{ChuDiffanly2020,ChenDiffnarrow2022}, the DMCANC system assumes that each node can receive information from other nodes in real time, which is unrealistic. In this case, we investigate the proposed DMCANC's tolerance to the network latency. Hence, we introduce a delay of time $p$ into (\ref{eq7}) and yield:
\begin{equation}\label{eq15}
    {w}_{k}(n) = {\psi}_{k}(n)-\sum_{m=1,m\neq k}^{N}{\psi}_{m}(n-p)*{c}_{km}(n).
\end{equation}
Figure~\ref{fig:7 Case2MSE} depicts the noise reduction performance of the proposed algorithm as $p$ increases. It demonstrates that the proposed DMCANC algorithm can successfully converge even when communication latency reaches $3000$ sampling periods, equivalent to $187.5$ ms when the system sampling frequency is set to $16000$ Hz. It is worth noting that the average WIFI network latency is less than $100$ ms. Therefore, the network communication in the proposed approach could be realized using a standard wireless communication protocol.
\begin{figure}[!t]
    \centering
    \includegraphics[width = 8cm,height=6cm]{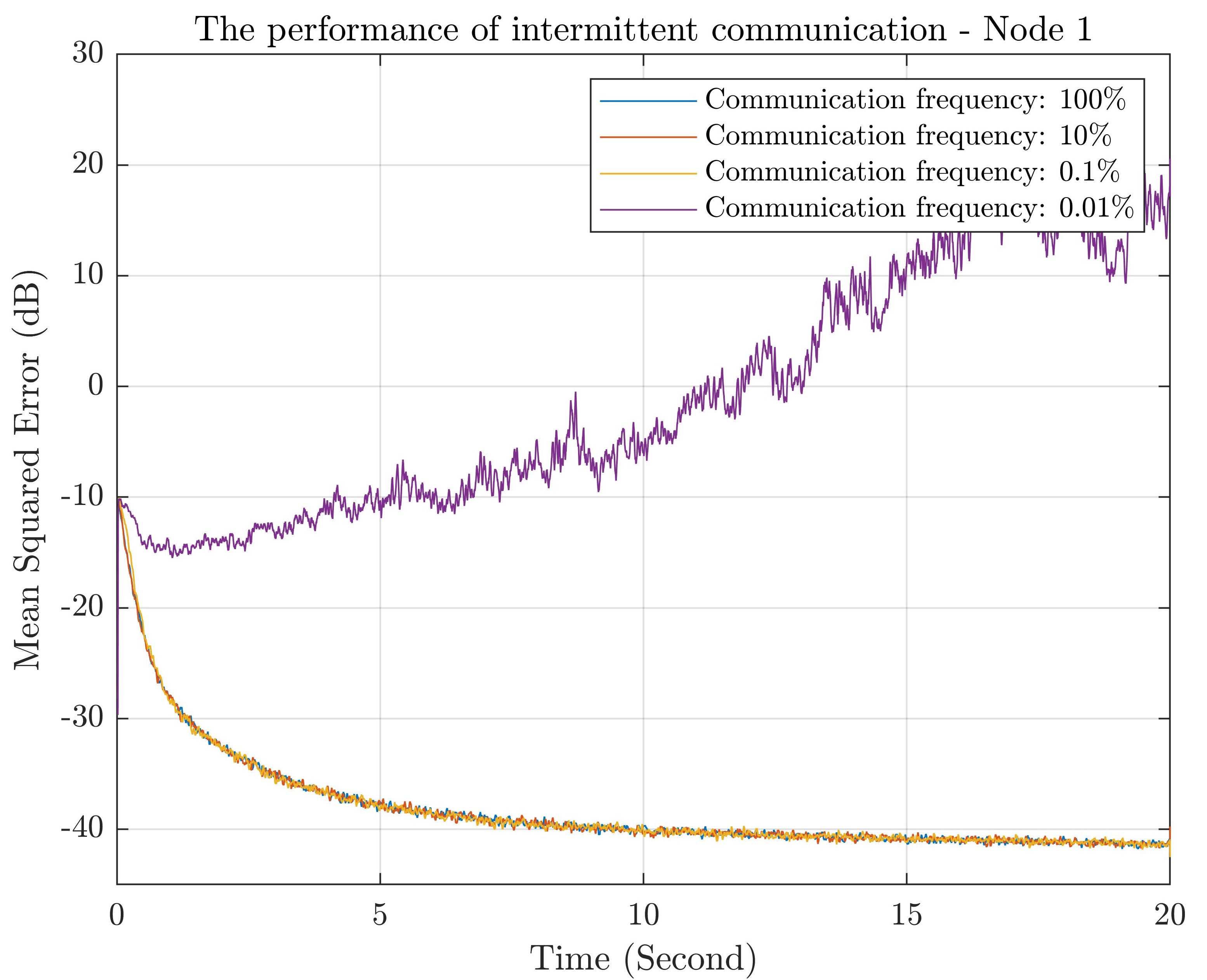}
    \caption{Mean Squared Error (MSE) of the proposed DMCANC with intermittent communication.}
    \label{fig:8 Case3MSE}
\end{figure}

{\it Simulation 3:} Since the proposed DMCANC can overcome the communication delay to some extent, communication interruptions are also a practical issue for most wired or wireless communication systems. To investigate how communication interruptions affect the performance of the proposed algorithm, we incorporate the intermittent information exchange mode into the proposed DMCANC algorithm, which can be expressed as:
\begin{equation}\label{eq16}
     {w}_{k}(n) =\left\{
     \begin{aligned}
        &{\psi}_{k}(n)-\sum_{m=1,m\neq k}^{N}{\psi}_{m}(n)*{c}_{km}(n), \quad &\beta_{k}=1,\\
        &{\psi}_{k}(n)-\sum_{m=1,m\neq k}^{N}{\psi}_{m}(n-q)*{c}_{km}(n), &\beta_{k}=0.
     \end{aligned}
\right.
\end{equation}
where $\beta_{k}=1$ represents that the $k$th node has communications with other nodes; $\beta_{k}=0$ shows no information exchange at the $k$th node and only uses the control filters received at the time ($n-q$). For a practical realization of the proposed DMCANC algorithm, we assume that the communication requests of each node are independent. In the simulation, we define communication frequency as the number of communication per second. The result in Fig.~\ref{fig:8 Case3MSE} illustrates that the proposed algorithm can accept lower communication frequency to a greater extent while achieving good MSE results. Even if the communication ratio is reduced to $0.1\%$, approximately 16 times per second, the system has a similar noise reduction performance as the one with uninterrupted communication. This indicates that the proposed algorithm can achieve good noise reduction performance even under a high incidence of communication interruptions.

\section{Conclusion}
\label{sec:conclusion}

This paper proposed a novel distributed multichannel active noise control (DMCANC) algorithm, which utilizes compensation filters to counteract cross-talk between nodes, while only updating the local control filter based on local error information. With these two mechanisms, it is possible to overcome practical communication issues, such as latency and interruptions, while maintaining steady-state optimal control performance. Accordingly, it is anticipated that this algorithm will be implemented in the distributed multichannel ANC system using standard wireless network protocols. In addition, the numerical simulation demonstrated a satisfactory noise reduction and a high degree of communication problem-solving stability.


\bibliographystyle{IEEEbib}
\bibliography{refs}

\end{document}